\documentclass[11pt]{article} 
\usepackage[utf8]{inputenc} 
\usepackage{geometry} 
\usepackage{graphicx} 

\usepackage{booktabs} 
\usepackage{array} 
\usepackage{paralist} 
\usepackage{verbatim} 
\usepackage{subfig} 
\usepackage{color}

\usepackage{fancyhdr} 
\usepackage{sectsty}
\allsectionsfont{\sffamily\mdseries\upshape} 

\usepackage[nottoc,notlof,notlot]{tocbibind} 
\usepackage[titles,subfigure]{tocloft} 

\title{Thermoelectric and thermomagnetic effects in Kaluza's kinetic theory}
\author{A. R. Sagaceta-Mejía$^{1}$ , A. Sandoval-Villalbazo$^{1}$ and A. L. García-Perciante$^{2}$ 
\bigskip
\\ $^{1}$ Departamento de Física y Matemáticas \\  Universidad Iberoamericana \\
Prolongación Paseo de la Reforma 880, Lomas de Santa Fe CP 01219 , Ciudad de México.\\
 e-mail:  alma.sagaceta@correo.uia.mx, alfredo.sandoval@ibero.mx
\smallskip
\\  $^{2}$ Departamento de Matemáticas Aplicadas y Sistemas \\ Universidad Autónoma Metropolitana-Cuajimalpa \\
Av. Vasco de Quiroga 4871, Cuajimalpa de Morelos, 05348 Ciudad de México.\\
e-mail: algarcia@correo.cua.uam.mx} 

\begin{document}
	\maketitle 
	\begin{abstract}
		A five-dimensional treatment of the Boltzmann equation is used to
		establish the constitutive equations that relate thermodynamic fluxes
		and forces up to first order in the gradients for simple charged fluids in the presence of electromagnetic fields.
		The formalism uses the \textit{ansatz} first introduced by Kaluza
		back in 1921, proposing that the particle charge-mass ratio is proportional
		to the fifth component of its velocity field. It is shown
		that in this approach, space-time curvature
		yields thermodynamic forces leading to generalizations of the well-known
		cross-effects present in linear irreversible thermodynamics.
	\end{abstract}
	
	\section{Introduction}
	
	According to the tenets of general relativity, classical fields can be considered
	consequences of space-time curvature produced by sources such as mass and electric charge. In the case of gravitational
	fields, Einstein's equations  show that the
	Newtonian potential can be interpreted as a metric coefficient, which
	in turn determines the geodesics that structureless particles  follow
	in physical situations such as planetary orbits. In the same spirit,
	electromagnetic potentials can generate similar effects curving space-time
	so that the trajectories of charged particles correspond to geodesics
	in a 5D space-time. 
	
    In Kaluza's approach \cite{kaluza1921sitzungsberichte}, the charge-mass ratio acts as the fifth component
	of the particle's velocity field; i.e.  $V^{5}=\frac{q}{m\xi}$ where $q$ is the charge, $m$ is the mass and   $\xi=\sqrt{16\pi G\epsilon_{0}/c^{2}}$ is the  coupling constant. The fundamental constants $G$, $\epsilon_0$ and $c$ have their usual meanings. Kaluza also proposed the cylindrical condition:
	 \begin{equation}\frac{\partial}{\partial  x^{5}}=0\label{eq:cil}\end{equation}
	 
	 \noindent which leads to the sourceless Maxwell equations and is compatible with the concept of periodic, compact extra dimensions \cite{kaluza1921sitzungsberichte,klein1926atomicity}. In
	 this work the capital latin letters run for 1 to 5, the Greek indices
	 will indicate values from 1 to 4, lowercase latin indices will
	 run from 1 to 3 and the dot corresponds to total proper time (arc length) derivatives. 
	
	\noindent In Kaluza's formalism,  the charged particles follow geodesics given by
	\begin{equation}
	\dot{V}^\mu=V^A V^{\mu} _{;A}=V^\alpha \frac{\partial V^{\mu}}{\partial x^\alpha}+\Gamma^\mu_{5B}V^B V^5=0\label{eq:lorentz}
	\end{equation}
	where $\Gamma^\mu_{AB}$ is the usual Christoffel symbol of the second kind:
	\begin{equation}
		\Gamma^\mu_{AB}=\frac{1}{2}\hat{g}^{\mu L}\left(\frac{\partial \hat{g}_{BL}}{\partial x^A}+\frac{\partial \hat{g}_{LA}}{\partial x^B}-\frac{\partial \hat{g}_{AB}}{\partial x^L}\right)
	\end{equation}
The metric considered contains the electromagnetic potential as follows,
	\begin{equation}
		\hat{g}_{AB}=\left[
		\begin{array}{ccccc}
		1&0&0&0&\xi \delta A_x\\
		0&1&0&0&\xi \delta A_y\\
		0&0&1&0&\xi \delta A_z\\
		0&0&0&-1&\frac{\xi}{c} \delta \phi\\
		\xi \delta A_x&\xi \delta A_y&\xi \delta A_z& \frac{\xi}{c} \delta \phi & 1\\
		\end{array}
		\right]
		\label{eq:gg}
	\end{equation}.
	Using Eqs. (\ref{eq:lorentz}-\ref{eq:gg}) one can immediately identify:
	\begin{equation}
	F_{\lambda}^{\mu}=-\frac{2}{\xi}\Gamma_{5\lambda}^{\mu}\label{eq:faraday}
	\end{equation}
	where $F_{\lambda}^{\mu}$ is the usual electromagnetic field tensor. Equation (\ref{eq:lorentz}) corresponds to the equation of motion of a charged particle in an electromagnetic field.
	\noindent 
	Maxwell's equations can be derived from Kaluza's ansatz and the tenets of general relativity. The connection between the electromagnetic force and the 5D geometry is given through Eq. (\ref{eq:faraday}). Following the standard approach, the potentials included in Eq. (\ref{eq:gg}) are here treated as perturbations of a flat space-time. All the second order terms in the calculations regarding these quantities will be assumed to be negligible.
 
   \noindent This 5D approach was first introduced to the phenomenological study of charged fluids within the framework of irreversible thermodynamics in Ref. \cite{sandovalPOP}. On the other hand, the foundations of the corresponding microscopic theory were only recently established in Ref. \cite{sandoval2015kinetic}. The present work revisits these ideas, including magnetic fields in a fully covariant fashion with the explicit introduction of the spatial projector which separates space and time components. This formalism, in which the field effects are consequences of the space-time curvature, is then used in order to obtain the heat flux and the electric current,  consistently  predicting the corresponding  effects related to transport processes in simple charged fluids.
	
	To accomplish this task, we have divided the paper as follows, in Section 2 we establish the  conservation laws using the Boltzmann equation in Kaluza's space-time. The  distribution function for the dissipative case to first order in the local gradients is obtained in Section 3 while the electromagnetic field contributions for the  heat flux and current density are established  in Section 4. A brief discussion of the results and final remarks are included in the last section.

	\section{Boltzmann equation}
	
	The Boltzmann equation for a simple dilute gas in the Kaluza space-time is expressed as
	\begin{equation}
		V^{A}f_{,A}+\dot{V}^{A}\frac{\partial f}{\partial V^{A}}=J(f,f^{'}),
		\label{eq:bolt}
	\end{equation}
	In Eq. (\ref{eq:bolt}) $J(f,f^{'})$ is the collisional kernel, $f$ is the distribution
	function and $V^{A}$ is the molecular five-velocity of a single particle, namely 
	\begin{equation}
		V^A=\left[
		\begin{array}{c}
		\gamma_{(v)}  v^\ell \\
		 c \gamma_{(v)}\\
		\frac{1}{\xi}\frac{q}{m}
		\end{array}
		\right]
	\end{equation}
	where $v^\ell$ is three-velocity and $\gamma_{(v)}$ is the usual Lorentz factor for the molecular velocity.

	\noindent In order to obtain the balance equations, we multiply Eq.  (\ref{eq:bolt}) by the collisional invariant $mV^B$ and integrate in velocity space \cite{garcia2011heat}. The conservation equations can be expressed in terms of  the particle and energy-momentum as follows (see the Appendix)
		\begin{equation}
	 T_{;B}^{AB}=0,
	\label{eq:tensor}
	\end{equation}
	where 
	\begin{equation}
	T^{AB}=m\int V^{A}V^{B}fd^{*}V
	\label{eq:inter}
	\end{equation}
	The invariant volume element can be expressed in terms of the peculiar velocity $k^\ell$ as \cite{garcia2011heat}
	\begin{equation}
	d^{*}V=\gamma^{5}c\frac{d^{3}k}{V^{4}}=\gamma^{4}d^{3}k,
	\label{eq:dV}
	\end{equation}
\noindent where $\gamma=\left( 1-\frac{k^\ell k_\ell}{c^2}\right)^{-\frac{1}{2}}$. It is important to notice that the charge (particle) four-flux is contained in Eq. (\ref{eq:inter}) using $A=5$ and $B=\mu$ namely,
	\begin{equation}
		J^\mu=\xi T^{5\mu}=\xi V^5 m\int f V^\mu d^{*}V=q\int f V^\mu d^{*}V=q N^\mu
\label{corriente}	
	\end{equation}

		\noindent Thus, in this approach particle conservation, given by $N_{;A}^{A}=0$, is included in Eq. (\ref{eq:inter}). 
		
		The Maxwell-Jüttner distribution
	function in equilibrium for relativistic particles is given by \cite{israel1963relativistic}
	\begin{equation}
	f^{(0)}=\frac{n}{4\pi c^{3}z\mathcal{K}_{2}\left(\frac{1}{z}\right)}e^{\frac{U_{\beta}V^{\beta}}{zc^{2}}}
   \label{eq:MJ}	
	\end{equation}
	where $\mathcal{K}_{2}\left(\frac{1}{z}\right)$ is the modified Bessel
	function of the second kind, $U_{\beta}$ the hydrodynamic four-velocity and $z=\frac{kT}{mc^2}$. Evaluation of Eq. (\ref{eq:inter}) considering the Jüttner distribution given in Eq. (\ref{eq:MJ}), leads to the following equilibrium particle and energy-momentum tensor:

	\begin{equation}
	T^{AB}=\left(\begin{array}{c|c}
	\frac{n\varepsilon}{c^{2}}U^{\mu}U^{\nu}+ph^{\mu\nu} & \frac{nq}{\xi}U^{\nu}\\
	\hline
	\frac{nq}{\xi}U^{\nu} & \frac{n}{\xi^{2}}\frac{q^{2}}{m^{2}} \frac{\mathcal{K}_{1}\left(\frac{1}{z}\right)}{\mathcal{K}_{2}\left(\frac{1}{z}\right)}
	\end{array}\right)\label{eq:10}
	\end{equation}

where the internal energy per particle is given by $\varepsilon=mc^2\left(3z+\frac{\mathcal{K}_{3}\left(\frac{1}{z}\right)}{\mathcal{K}_{2}\left(\frac{1}{z}\right)}\right)$. Substitution of $T^{5\mu}$ from Eq. (\ref{eq:10}) in Eq. (\ref{eq:inter}) leads to the continuity equation namely, 
\begin{equation}
\dot{n}+n\theta=0\label{eq:12}
\end{equation}
where we have defined the total derivative as $\dot{n}=U^{A}n_{,A}$ and
$\theta=U_{;A}^{A}$. Notice that due to the cylindrical condition given by Eq. (\ref{eq:cil}), the particle conservation equation is identical to the one obtained in the four dimensional formalism.

	The internal energy balance follows from projecting Eq. (\ref{eq:tensor}) in the direction of the hydrodynamic four-velocity, which leads to \cite{sagaceta2016statistical}
	
	\begin{equation}
	n\dot{\varepsilon}+p\theta=0\label{eq:17}
	\end{equation}
	On the other hand, Euler's equation for the evolution of the hydrodynamic four-velocity is obtained by explicitly calculating the covariant derivative of the energy-momentum tensor, which yields \cite{sagaceta2016statistical}

\begin{equation}
\tilde{\rho}\dot{U}^{\nu}+h^{\nu\alpha}p_{,\alpha}=-2\Gamma_{\alpha 5}^{\nu}T^{5\alpha}\label{eq:8}
\end{equation}

\noindent The Christoffel symbol included in Eq. (\ref{eq:8}) is
easily identified with the electromagnetic field tensor in accordance
to Eq. (\ref{eq:faraday}), so that the well known relativistic hydrodynamic
momentum balance equation in the Euler regime is recovered:

\begin{equation}
\tilde{\rho}\dot{U}^{\nu}+h^{\nu\alpha}p_{,\alpha}=nqF_{\lambda}^{\nu}U^\lambda
\label{eq:19}
\end{equation}
where $\tilde{\rho}$ is given by 
	
	\begin{equation}
	\tilde{\rho}=\left(\frac{n\varepsilon}{c^{2}}+\frac{p}{c^{2}}\right)\label{eq:21}
	\end{equation}
	The right hand side of Eq. (\ref{eq:19}) corresponds to the Lorentz force which can be thus viewed
	as an effect of space-time curvature. We are now in position to analyze
	the dissipative effects up to first order in the gradients.
	
	 \section{Navier-Stokes regime}
	Dissipative effects, which deviate the distribution function from its local equilibrium value, can be introduced through the Chapman-Enskog expansion \cite{chapman} namely,
	\begin{equation}
	f=f^{(0)}+\mu f^{(1)}+\mu^{2}f^{(2)}+\ldots
	\end{equation}
	where the parameter $\mu$ is the Knudsen number.
	
	\noindent	For the sake of simplicity,  the right hand side of the Boltzmann equation is here modeled using a relaxation approximation as follows
	\begin{equation}
	J(f,f^{'})=-\frac{f-f^{(0)}}{\tau_c}
	\label{eq:BGK}
	\end{equation}
Use of the cylindrical condition (\ref{eq:cil}) and the metric (\ref{eq:gg}) in the standard treatment of the Boltzmann equation within the BGK type model here introduced yields 
	
	\begin{equation}
	f^{\left(1\right)}=-\tau_c
	\left[ 
	V^{\alpha}\left(\frac{\partial f^{\left(0\right)}}{\partial n}\frac{\partial n}{\partial x^{\alpha}}+\frac{\partial f^{\left(0\right)}}{\partial T}\frac{\partial T}{\partial x^{\alpha}}+\frac{\partial f^{\left(0\right)}}{\partial U^{B}}U_{;\alpha}^{B}\right)
	+ \dot{V}^A \frac{\partial f^{(0)}}{\partial V^A}	\right]
	\label{eq:23}
	\end{equation}
	
	\noindent It is important to notice that the last term of Eq. (\ref{eq:23}) vanishes since particles move following  geodesics, $\dot{V}^A=0$, and thus 
	
	\begin{equation}
	f^{\left(1\right)}=-\tau_c V^{\alpha}\left(\frac{\partial f^{\left(0\right)}}{\partial n}\frac{\partial n}{\partial x^{\alpha}}+\frac{\partial f^{\left(0\right)}}{\partial T}\frac{\partial T}{\partial x^{\alpha}}+\frac{\partial f^{\left(0\right)}}{\partial U^{\beta}}U_{;\alpha}^{\beta}\right)
	\label{eq:23-1}
	\end{equation}
		
	\noindent The index $\beta$ in Eq. (\ref{eq:23-1}) appears instead  of $B$ in Eq. (\ref{eq:23})  since $f^{(0)}$ is independent of the electric charge. The first two terms of Eq. (\ref{eq:23-1})  have been already discussed in the literature of relativistic fluids and lead to the coupling of the dissipative fluxes with the temperature and density gradients \cite{cercignani2002relativistic}. Here we focus our attention in the last
term, which contains the electromagnetic field and can be written as
	\begin{equation}
	V^{\alpha}\frac{\partial f^{\left(0\right)}}{\partial U^{\beta}}U_{;\alpha}^{\beta}=\frac{m}{kT}V^{\alpha}V_{\beta}\left(\frac{\partial U^{\beta}}{\partial x^{\alpha}}+\Gamma_{\alpha L}^{\beta}U^{L}\right)f^{\left(0\right)}=\frac{m}{kT}V^{\alpha}V_{\beta}\left( U^{\beta}_{;\alpha}+\Gamma_{\alpha 5}^{\beta}U^{5}\right)f^{\left(0\right)}
	\label{eq:24}
	\end{equation}
where the second term after the last equality vanishes since it contains the contraction of a symmetric and an antisymmetric tensor. Also, for the remaining term in Eq. (\ref{eq:24}) it is convenient to separate space and time components using Eckart's decomposition as follows
	\begin{equation}
		V^{\alpha}=\gamma U^{\alpha}+\mathcal{R}_{\lambda}^{\alpha}K^{\lambda} \label{eq:25}
	\end{equation}
here $\mathcal{R}_{\lambda}^{\alpha}=h_{\nu}^{\alpha}L_{\lambda}^{\nu}$, $h^\alpha_\mu=\delta^\alpha_\mu+\frac{1}{c^2}U^\alpha U_\mu$,  $L_\lambda^\alpha$ is the Lorentz boost and $K^\mu=\gamma\left[k^\ell,c\right]$ \cite{eckart1940thermodynamics, garcia2012microscopic}. Following the Chapman-Enskog method, and in order to guarantee existence of the solution, the time derivatives of the state variables are written in terms of the spatial gradients by introducing the Euler equations. This step in the procedure leads to the coupling of the non-equilibrium distribution function with the gradient of the electrostatic potential. Indeed, the introduction of Eq. (\ref{eq:25}) in  Eq. (\ref{eq:24}) leads to the following expression

	\begin{equation}
\frac{m}{KT}f^{(0)}\mathcal{R}^\mu_\beta K_\mu \gamma \dot{U}^\beta=\frac{m}{\tilde{\rho}kT}f^{(0)}\mathcal{R}^\mu_\beta K_\mu \gamma \left(nqU_\eta F^{\beta\eta}-h^{\beta\eta}p_{,\eta}\right)
	\label{eq:26}
	\end{equation}
where we have used Euler's equation (\ref{eq:19}). It is natural to define the electromagnetic contribution to the non-equilibrium distribution function as follows
\begin{equation}
f^{(1)}_{[EM]}=-\tau_c \gamma\frac{q}{kT}\frac{nm}{\tilde{\rho}}\mathcal{R}^\mu_\beta K_\mu  U_\eta F^{\beta\eta}f^{(0)}
\label{eq:fiE}
\end{equation}
which is similar to the corresponding  expression established in Ref. \cite{garcia2015relativistic} by using special relativistic arguments. However, in the present case the contribution does not vanish in the low temperature limit and thus Eq. (\ref{eq:fiE}) can be used to calculate the current density in such regime. This will be addressed in the next section.
	

	\section{Electric current and heat flux}
	In order to obtain an expression for the electric current, the first order correction to the distribution function due to the electromagnetic field given in Eq. (\ref{eq:fiE}) is introduced in the definition (\ref{corriente}):
	
	\begin{equation}
	J_{[EM]}^{\nu}=-\tau_c \frac{q^2}{kT}\frac{n m}{\tilde{\rho}}\mathcal{R}^\mu_\beta \mathcal{R}^\nu_\lambda U_\alpha F^{\alpha \beta}\int \gamma^3 k^{\lambda} k_{\mu} f^{(0)} d^{*}K
\label{eq:J1}	
	\end{equation}.

	\noindent Here the volume element is  $d^{*}K=4\pi c^{3}\sqrt{\gamma^{2}-1}d\gamma$ (see Appendix B of Ref. \cite{garcia2011heat}). The integral in Eq. (\ref{eq:J1}) can be directly evaluated and leads to the following expression for the electrical current

		\begin{equation}
	J^\nu_{[EM]}=-\tau_c \frac{n q^2}{m}\mathcal{R}^\lambda_\beta \mathcal{R}^\nu_\lambda U_\alpha F^{\alpha \beta} 
	\label{eq:electrical}
	\end{equation}

	\noindent where use has been made of the fact that $\tilde{\rho}=nm\mathcal{G}\left(\frac{1}{z}\right)$, with $\mathcal{G}\left(\frac{1}{z}\right)=\frac{K_3\left(\frac{1}{z}\right)}{K_2\left(\frac{1}{z}\right)}$. Notice that in the electrostatic case $J^\ell =\sigma E^\ell$ where $\sigma$ is easily identifed as the electrical conductivity, which in the non-relativistic case is given by $\sigma=\frac{nq^2}{m}\tau_c$. This expression is similar to its non-relativistic counterpart, obtained using the stationary state approximation \cite{mckelvey1966solid}. It is important to emphasize at this point that the relation between the electrical current and the electromagnetic field cannot be directly established following standard kinetic theory without neglecting the partial time derivative term in Boltzmann's equation or introducing similar assumptions. The result in Eq. (\ref{eq:electrical}), which is the electromagnetic generalization of the one obtained in Ref. \cite{sandoval2015kinetic}, shows that the introduction of the electromagnetic potential in the space-time geometry allows for Ohm's law to be established from microscopic grounds without such assumptions. Moreover, the dependence of the corresponding conductivity with the temperature, here given through the relaxation time $\tau_c$, is consistent with the theoretical predictions which have been thoroughly verified.
	
	The present formalism also allows to compute the electromagnetic contribution to the heat flux considering the definition given in Ref. \cite{garcia2012microscopic} namely,
	
	\begin{equation}
	J_{[Q, EM]}^{\nu}=mc^{2} \int \left(\gamma-1\right)\gamma K^\nu f^{(1)}_{[EM]}d^{*}K
	\end{equation}
which yields
 
\begin{equation}
	J_{[Q,EM]}^{\nu} = -\tau_c n q m c^{2}\left\{5z-1+\frac{1}{\mathcal{G}\left(\frac{1}{z}\right)}\right\}U_{\eta}F^{\alpha\eta}\mathcal{R}_{\beta}^{\lambda}\mathcal{R}_{\lambda}^{\nu}
	\label{eq:heatel}
\end{equation}
In the electrostatic case, we can identify the relativistic Benedicks coefficient as 

\begin{equation}
\mathcal{B}=\tau_c n k c^2 \left\{5z-1+\frac{1}{\mathcal{G}\left(\frac{1}{z}\right)}\right\}
\end{equation}
such that the corresponding constitutive equation reads
\begin{equation}
J^\ell_{[Q,E]}=-\mathcal{B}\frac{q m}{k}E^\ell
\end{equation}
Moreover, considering small values of $z$ one has 
$\mathcal{B}\sim\frac{5}{2 m}n k^2 T \tau_c$ which leads to a Wiedemann-Franz type relation namely,

\begin{equation}
	\frac{\mathcal{B}}{\sigma}=\frac{5}{2}\frac{k^2}{q^2}T
\end{equation}

This result is independent of $n$ and $m$ as in the case of free electrons in metals. The original definition of this relation is given through the classical Lorenz number $\mathcal{L}=\frac{5}{2}\frac{k^2}{q^2}$, as the ratio of the thermal and electric conductivities \cite{kittel1966introduction}, i.e. 

\begin{equation}
	\frac{k_{th}}{\sigma}= \mathcal{L}T
\end{equation}

	\section{Final Remarks}
	Equations (\ref{eq:electrical}) and (\ref{eq:heatel}) are the main results of this work. It is important to notice that both expressions arise from the contribution of the fluid as a whole, through the introduction of Euler's equation in the expression for $f^{(1)}$, as part of the Chapman-Enskog procedure. This contribution, associated with the bulk, is not present in the stationary state approach since the partial time derivative term of the equilibrium distribution function is neglected. Space-time curvature is identified as the source driving the dissipative fluxes through the covariant derivatives present in Kaluza's-type approach.
	
	The present formalism predicts stronger thermoelectric and thermomagnetic effects than the ones obtained in the special relativistic approach (see Eq. (32) in \cite{garcia2013benedicks} and Eq. (43) in \cite{garcia2015relativistic}). Indeed the coefficient here obtained is independent of the speed of light in the low temperature limit. These effects are measurable at low temperatures ($z\ll 1$) and are described in the theory of free electrons in solid state physics text books within the stationary state approximation \cite{kittel1966introduction, mckelvey1966solid}. Also, the fact that Ohm's law is recovered constitutes an improvement with respect to the 4-dimensional formalism. The present paper generalizes such result, already established for the electrostatic case in Ref. \cite{sandoval2015kinetic}, to the electromagnetic case in an arbitrary reference frame.

	\noindent Future work includes the establishment of the complete set of transport equations including the general force-flux relations in the dynamics of the charged fluid. The stability analysis of such equations  under linear perturbations in the state variables and the establishment of the corresponding entropy production for the simple charged fluid in a curved space-time are also important problems to be addressed. The present formalism will also be generalized to a binary mixture in order to obtain the contribution to the cross-effects and the corresponding reciprocity relations. The approach here presented is promising and could lead to the identification of unnoticed general relativistic-type effects in the physics of charged fluids.
	
\section*{Acknowledgements}
The authors wish to thank Dominique Brun-Battistini for her valuable comments to this work.
	
\section*{Appendix A}
	In this appendix we establish the conservation relations given by Eq. (\ref{eq:tensor}) in a 5D space-time. The starting point is the Boltzmann equation given by 
	\begin{equation}
	V^{A}f_{,A}=J\left(ff^{'}\right)
	\end{equation}
Multiplying both sides by the collisional invariant $m V^B$, and integrating with respect to the velocity space volume element, leads to
	\begin{equation}
	m\int V^{A}V^{B}f_{,A}d^{*}V =0
	\label{eq:in1}
	\end{equation}
In order to derive Eq. (\ref{eq:tensor}), we make use of the identity $$	\left(V^{A}V^{B}f\right)_{;A}=V^{A}V^{B}f_{,A}+\left(V^{A}V^{B}\right)_{;A}f$$ such that Eq. (\ref{eq:in1}) reads

		\begin{equation}
	m\int V^{A}V^{B}f_{,A}d^{*}V   =\left(m\int V^{A}V^{B}fd^{*}V\right)_{;A}-m\int f\left(V^{A}V^{B}\right)_{;A}
	\label{eq:in}
	\end{equation}
The first term on the right hand side is readily identified as $T^{AB}_{;A}$. For the second term we use 
	
		\begin{equation}
	\left(V^{A}V^{B}\right)_{;A} =V_{;A}^{A}V^{B}+V^{A}V_{;A}^{B}
	=V_{;A}^{A}V^{B}+\dot{V}^{B}
	\end{equation}
where  $\dot{V}^{B}=0$, since particles move along geodetic paths. The covariant derivative term $V_{;A}^{A}$ also vanishes since space and velocity variables are mutually independent and thus

\begin{equation}	
	V_{;A}^{A}  =\Gamma_{AL}^{A}V^{L}
	\end{equation}
which can be shown to be of second order, and thus negligible, in Kaluza's classical formalism by direct calculation of the Christoffel symbols using the metric tensor given in Eq. (\ref{eq:gg}). Using this fact in Eq.  (\ref{eq:in}) leads directly to expression  (\ref{eq:tensor}) in the main text.
	

	

\end{document}